\documentclass{article}
\usepackage{graphicx}  
\usepackage{amsmath}   
\usepackage{amssymb}   
\usepackage[export]{adjustbox}
\usepackage{bm} 
\usepackage{dcolumn}
\usepackage{color}
\usepackage{mathrsfs}
\usepackage{amsfonts}
\usepackage{varioref}
\RequirePackage[colorlinks,citecolor=blue,urlcolor=magenta,linkcolor=blue]{hyperref}
\labelformat{section}{Section #1} 
\labelformat{subsection}{Section #1} 
\labelformat{subsubsection}{Section #1}
\labelformat{subsubsubsection}{Section #1}
\labelformat{equation}{Eq.~(#1)} 
\labelformat{figure}{Fig.~#1} 
\labelformat{subfigure}{Fig.~\thefigure#1} 
\labelformat{table}{Tab.~#1} 
\labelformat{appendix}{Appendix #1}

\title{\bf Constraining phantom braneworld model from cosmic structure sizes}
\author{\bf Sourav Bhattacharya$^{1}$\footnote{sbhatta@iitrpr.ac.in}~~and~ Stefanos R Kousvos$^{2}$\footnote{skousvos@physics.uoc.gr} \\
$^{1}$\small{Department of Physics, Indian Institute of Technology Ropar, Rupnagar, Punjab 140 001, India}\\
$^{2}$\small{ITCP and Department of Physics, University of Crete, 700 13 Heraklion, Greece} }

\begin{document}
  
\maketitle
\begin{abstract}
\noindent
We consider the  phantom braneworld model in the context of the maximum turn around radius, $R_{\rm TA,max}$, of a stable, spherical cosmic structure with a given mass. The maximum turn around radius is the point where the attraction due to the central inhomogeneity gets balanced with the repulsion of the  ambient dark energy, beyond which a structure cannot hold any mass, thereby giving the maximum upper bound on the size of a stable structure. In this work we  derive an analytical expression of $R_{\rm TA,max}$ for this model using cosmological scalar perturbation theory. Using this we numerically constrain the  parameter space, including a bulk cosmological constant and the Weyl fluid, from the mass versus observed size data for some nearby, non-virial cosmic structures. We use different values of the matter density parameter $\Omega_m$, both larger and smaller than that of the $\Lambda{\rm CDM}$, as the input in our analysis. We show in particular,  that  a) with a vanishing bulk cosmological constant the predicted upper bound is always greater than what is actually observed; similar conclusion holds if the bulk cosmological constant is negative b) if it is positive,  the predicted maximum size can go considerably below than what is actually observed and owing to the involved nature of the field equations, it leads to interesting constraints on not only the bulk cosmological constant itself but on the whole parameter space of the theory.  
\end{abstract}
\vskip .5cm
\noindent
\noindent
{\bf keywords :} Braneworld model, large scale structures, maximum turn around radius 
\bigskip
\section{Introduction}\label{sec.1}
\noindent
The braneworld model is a rather radical proposal in the effort to understand the large scale structure of the universe -- in which the $(3+1)$ dimensional universe we live in, is just a timelike hypersurface (the brane) of codimension one or more, embedded in a higher dimensional spacetime (the world), see~\cite{Brax:2004xh, Maartens:2010ar}  for a vast review and also references therein. In this picture, unlike the higher dimensional theory like the Gauss-Bonnet gravity (e.g.~\cite{Maeda:2006hj} and references therein),  {\it all} standard model matter fields are confined on the brane whereas only  the gravitons can propagate in the extra dimension(s).  The braneworld model could be hoped, at least from the perspective of the string theory, to be a  possible bridge between Einstein's theory of gravitation and a hitherto unknown complete theory of quantum gravity.

Such shift of paradigm certainly is a modification of gravity and for example in the celebrated Randall-Sundrum model~\cite{Randall:1999ee, Randall:1999vf} with brane-codimension one, the modification is in the small scale. The extra dimension need neither be small nor even compact -- in fact it could be large compared to the fundamental length scale observed on the brane and can even be infinite. Compact extra dimensions, on the other hand, can give rise to an infinite and discrete tower of graviton mass when viewed on the brane, see e.g.~\cite{Durrer:2003rg}. We further refer our reader to e.g.~\cite{Mak:2004hv}-\cite{Harko:2007yq}  for an account on fitting the galaxy rotation curves and derivation of gravitational lensing, in the context of the Randall-Sundrum model.  While the extra dimension along the brane is usually taken to be spacelike, we refer our reader to~\cite{Shtanov:2002mb} for an interesting discussion on timelike extra dimension.

 In the front of gravity oriented computations, discussions on stationary black holes or other static solutions can be seen in e.g.~\cite{Kanti:2013lca, Kanti:2015poa, Nakonieczna:2016qlm, Kar:2015fva} and also references therein.  See~\cite{Yagi:2016jml} for a review on the constraints on such models based on gravity oriented tests. In particular, for the so called two branch RS-I model, from the modification of  Newton's law, the upper bound on the bulk anti-de Sitter radius turns out to be $l\lesssim 14\, \mu m$; whereas for the one branch RS-II model, the binary gravity wave data puts a bound : $l\lesssim 3.9\, \mu m$.    Probing the extra dimensional effects by studying the strong gravitational lensing can be seen in~\cite{Chakraborty:2016lxo}. We further refer our reader to~\cite{CliftonEtAl2012} for a vast review and an exhaustive list of references pertaining gravity and cosmology in the context of the braneworld model.

An important class of braneworld model which we shall be concerned with in this paper, is the Dvali-Gabadadze-Porrati braneworld (DGP) model~\cite{Collins:2000yb}-\cite{Shtanov:2000vr} containing in the action,  the 4-dimensional Ricci scalar on the brane, induced from the one loop correction due to the graviton-matter interaction and as well the extrinsic curvature of the brane. This class of models can significantly modify gravity at large scales unlike the Randall-Sundrum model. It gives two branches of cosmological solutions, both with flat spatial sections, one being self accelerated without requiring any dark energy/cosmological constant, whereas the other branch (the normal branch) requires at least one cosmological constant to make the current accelerated expansion~\cite{Deffayet:2000uy, Deffayet:2001pu, Charmousis:2006pn}.  Unfortunately, the former was shown to have ghost instability in later works~\cite{Gorbunov:2005zk, Koyama:2007zz}, leaving the normal branch with a cosmological constant to be an alternative to $\Lambda{\rm CDM}$.

For DGP models, the effective equation of state parameter $w$ ($P=w\rho$), is time dependent  and turns out to be less than minus of unity for the current cosmology~\cite{Sahni:2002dx}-\cite{Shtanov:2005}. However,  subject to parameter values, $w(t)$ could asymptotically reach $-1$ (the de Sitter phase) or even the universe  can leave at some stage the phase of accelerated expansion reentering matter domination, eventually thus evading the phantom disaster~\cite{Weinberg:2008zzc} for $w<-1$ dark energy models in the General Relativity. Such models, for $w$ being less than minus of unity in the current cosmological epoch, are known as the phantom braneworld. Interestingly, this model indicates that the expansion of our universe was stopped at redshift $z\gtrsim 6$ and `loitered' there for a long period of time encouraging the structure formation. This seems to be in agreement with the observed population of the quasistellar objects and supermassive black holes in
$6\lesssim z\lesssim 20 $~\cite{Alam:2005pb}.  Discussion of scalar cosmological perturbation theory in the phantom braneworld model and a more detailed review  can be seen in~\cite{Viznyuk:2012oda, Bag:2016tvc}.

The $\Lambda{\rm CDM}$  has its most remarkable simplicity and overwhelming phenomenological successes, starting from the type Ia supernovae redshift, the structure formation, the anisotropy in the primordial radiation and so on, e.g.~\cite{Weinberg:2008zzc} and references therein\footnote{See however, also~\cite{sarkar} for a recent critique to the data that led to infer the accelerated cosmic expansion.}. However, the lack of any satisfactory theory to explain the current tiny observed value of $\Lambda$ (the so called fine tuning problem of $\Lambda$~\cite{Martin2012}), lack of any quantum field theoretic mechanism to explain how it reached such a tiny value starting from an initial higher value during inflation (the so called coincidence problem~\cite{Tomaras1986, Floratos1987, Polyakov2012, Woodard2014}) and finally, the so far lack of any evidence of a dark matter particle, have triggered vigorous researches in  alternatives to the $\Lambda{\rm CDM}$ model. We refer our reader to~\cite{CliftonEtAl2012} for a vast review and an exhaustive list of references in various alternative gravity/dark energy models. We further refer the reader to~\cite{Sahni2014}, where from the  baryon acoustic oscillation data, it has been argued that the Hubble rate at high redshift should actually be less than the prediction of  $\Lambda{\rm CDM}$.

A couple of years back, a very novel check of different dark energy models was proposed in~\cite{PavlidouTomaras2013, PavlidouTetradisTomaras2014}, using the idea of the maximum turn around radius of some nearby cosmic structures. The maximum turn around radius $R_{\rm TA,max}$ is basically the point where the attraction on a radially moving test particle/fluid due to the central inhomogeneity gets precisely balanced with the repulsion due to the ambient dark energy. For $\Lambda{\rm CDM}$, one obtains $R_{\rm TA, max}=(3MG/ \Lambda)^{1/3}$, for spherical structures. See~\cite{Stuclik1} for a discussion on the maximum turn around radius in the context of the Schwarzschild-de Sitter black hole and also~\cite{Eingorn:2012dg} for a derivation alternative to that of~\cite{PavlidouTomaras2013, PavlidouTetradisTomaras2014}.   Clearly, a structure cannot hold any mass  beyond $R_{\rm TA, max}$  and thus a stable structure can never be larger than that. In other words, if a certain dark energy/gravity theory predicts a structure size smaller than what is actually observed, it gets ruled out.

The actual observed sizes of some nearby $(z \ll 1)$ cosmic structures was compared with the theoretical prediction of $\Lambda{\rm CDM}$ in~\cite{PavlidouTomaras2013, PavlidouTetradisTomaras2014}, and was shown that for those with masses $M \gtrsim 10^{14}M_{\odot}$, this theoretical upper bound is only roughly about $10 \%$ greater than  the observed sizes. The result was also generalized for dark energy with equation of state $P(t)=w\rho(t)$ with a constant  $w$, to get a strict bound $w>-2.3$. In~\cite{TanoglidisPavlidouTomaras2014, TanoglidisPavlidouTomaras2016}, it was shown using the Press-Schechter statistical mass function that structures  
with $M\gtrsim 10^{13}M_{\odot} $ are {\it not} yet virialized today and hence larger in size than the virialized ones, and are close to the theoretical maximum upper bound. We refer our reader to~\cite{Lee:2015upn, Lee2016, Lee:2016oyu} for discussions on looking for violation of the maximum turn around bound via the peculiar velocity profile of the constituents of structures. Note that contrary to other conventional tests of the dark energy (e.g. redshift of the distant type Ia supernovae), this is completely a local check , as the structures we observe are sufficiently nearby $(z \ll 1)$ -- in fact it is a demonstration of the fact that the dark energy is at work right within the structures.
  
 The maximum turn around perspective has received considerable attention in the modified gravity sector. It has been applied to theories like the Brans-Dicke, galileon, quintessence and the generalized Chaplygin gas models, a generic dark energy with completely arbitrary state `parameter' $w(t)$ and so on~\cite{Faraoni2015}-\cite{Bhattacharya:2017yix}. In particular, a general derivation of the maximum turn around radius for theories satisfying the Einstein equivalence principle can be seen in~\cite{Bhattacharya:2016vur}. From all these follow up works of~\cite{PavlidouTomaras2013, PavlidouTetradisTomaras2014} so far,  it has been well established  to use  $R_{\rm TA, max}$ as an observable for the aforementioned non-virial cosmic structures to constrain dark energy/alternative gravity  models. We further refer our reader to~\cite{Bhattacharya:2017yix} and references therein for an elaborate review with a list of references. See also~\cite{Merafina:2014ysa, Donnari:2016pmd} for a different proposal for a local check of the dark energy, from the observation of motion of different galaxies. 

The braneworld scenario,  is certainly {\it qualitatively different } from the examples we have given above, for additional effects that stem from the extra dimension(s). Being motivated by this, we shall derive in~\ref{sec.3} the maximum turn around radius for the DGP model considered in~\cite{Bag:2016tvc} and references therein. We shall show that if we set the bulk cosmological constant to zero, the maximum turn around radius predicted is always larger than the $\Lambda{\rm CDM}$. However, interesting things happen if we turn on the bulk cosmological constant  $\Lambda$ -- due to the nontrivial field equations, it can  actually `interact' with the cold dark matter and the brane cosmological constant leading to interesting bounds on the whole parameter space of the theory,~\ref{sec.3.1}. Finally we conclude in~\ref{sec.4}.

We shall work with the mostly positive signature of the metric $(-,+,+,+,+)$ and will set $c=1$.

\section{The model and the field equations}\label{sec.2}
\noindent
In the following we shall briefly review the action and the field equations for the DGP model, the details of which can be seen in e.g.~\cite{Shtanov:2005}.  The system is described by the action~\cite{Shtanov:2000vr, Shtanov:2005, Bag:2016tvc},
\begin{equation}
S=M^3 \left[\int_{\rm Bulk}(\mathcal{R}-2\Lambda)-2\int_{\rm Brane}K\right]+\int_{\rm Brane}(m^2 R-2\sigma)+\int_{\rm Brane} {\cal L}(g_{\mu\nu},\phi)
\label{bw1}
\end{equation}
The brane has been taken to have codimension one.  $M$ and $m$ are respectively the $5$ and $4$ dimensional Planck masses whereas $\mathcal{R}$ and $R $ are the corresponding Ricci scalars. $\Lambda$ is the cosmological constant in the bulk, $K$ is the extrinsic curvature on the brane, the combination of $\sigma/m^2$ plays the role of the cosmological constant on the brane with $ \sigma$ being the brane tension.  ${\cal L}$ stands for the Lagrangian density of matter fields living  on the brane, for our current purpose which would be the cold dark matter.

Variation of the above action leads to the field equations projected onto the brane, e.g.~\cite{Shtanov:2000vr},
\begin{equation}
G_{\mu\nu} + \left(\frac{\Lambda_{\rm RS}}{b+1}\right)g_{\mu\nu}=\left(\frac{b}{b+1}\right)\frac{1}{m^2}T_{\mu\nu}+\left(\frac{1}{b+1}\right)\left[\frac{1}{M^6}Q_{\mu\nu}-\mathcal{C}_{\mu\nu}\right]
\label{bw2}
\end{equation}
where the parameters entering this equation are defined as
\begin{equation}
b=\frac{\sigma l}{3M^3} \qquad l=\frac{2m^2}{M^3} \qquad \Lambda_{\rm RS}=\frac{\Lambda}{2}+\frac{\sigma^2}{3M^6}
\label{bw3}
\end{equation}
and
\begin{equation}
Q_{\mu\nu}=\frac{1}{3}EE_{\mu\nu}-E_{\mu\lambda}E_{\nu}^\lambda +\frac{1}{2}\left (E_{\rho\lambda}E^{\rho\lambda}-\frac{E^2}{3} \right)g_{\mu\nu}
\label{bw4}
\end{equation}
where
\begin{equation}
E_{\mu\nu}=m^2G_{\mu\nu}-T_{\mu\nu} \quad {\rm and } \quad
E=E^\mu{}_\mu
\label{bw5}
\end{equation}
and ${\cal C}_{\mu \nu} $ comes from the projection of  the bulk Weyl tensor onto the brane.
We take the Friedman-Robertson-Walker (FRW) ansatz for the metric with flat spatial sections and conformal time
\begin{equation}
ds^2= a^2(\tau)\left[ -d\tau^2+ dx^2+dy^2+dz^2 \right]
\label{bw6}
\end{equation}
and plug it into \ref{bw2}, with $T_{\mu \nu}$ corresponding to the cold dark matter; the contribution coming from the Weyl part  is seen to behave like the electromagnetic radiation density $\sim 1/a^4(\tau)$  and we shall ignore it from the homogeneous cosmological equations. However, we shall consider its inhomogeneous contribution later. One then obtains, for the so called normal branch, 
\begin{equation}
\mathcal{H}^2=\frac{a^2\left(\rho+\sigma\right)}{3m^2}+\frac{2a^2}{l^2} \left[1-\sqrt{1+l^2\left(\frac{\rho+\sigma}{3m^2}-\frac{\Lambda}{6} \right)}\right] 
\label{bw7}
\end{equation}
where $\rho$ stands for the   cold dark matter energy density, ${\cal H}=\dot{a}/a$ is the Hubble rate, where a `dot' stands for derivative once with respect to $\tau$. Initially we set the bulk cosmological constant $\Lambda$ to zero, to get 
\begin{equation}
\mathcal{H}^2=\frac{a^2}{l^2}\left[\sqrt{1+l^2\left(\frac{\rho+\sigma}{3m^2}\right)}-1 \right]^2
\label{fried}
\end{equation}
It is customary then to define two useful quantities, $\beta$ and $\gamma$, as
\begin{equation}
\beta=-2\sqrt{1+l^2\left(\frac{\rho+\sigma}{3m^2}\right)}=-2\left(1+\frac{l \mathcal{H}}{a}\right) \qquad
3\gamma-1=\frac{ \dot{\beta}}{{\mathcal{H}}\beta}=\frac{\partial_{\tau}({\mathcal{H}}/a)}{{\mathcal{H}}\left(1+l \mathcal{H}/a\right)}
\label{bw8}
\end{equation}
With the help of the cosmological density functions : $\Omega_m(a)=\rho a^2 /(3m^2 {\mathcal H}^2)$, $\Omega_\sigma(a)=\sigma a^2/(3m^2 {\mathcal{H}}^2) $ and $\Omega_l(a)=a^2/(l^2 \mathcal{H}^2)$, and recalling that  $\rho \sim a^{-3}(\tau)$ for the cold dark matter, the above quantities can be re-expressed as 
\begin{equation}
\beta=-\frac{2}{\sqrt{\Omega_l}} \sqrt{\Omega_m(1+z)^3+\Omega_\sigma+ \Omega_l} \qquad
3\gamma-1=-\frac{3\Omega_m(1+z)^3}{2\left(\Omega_m(1+z)^3+\Omega_\sigma+\Omega_l\right)}
\label{bw9}
\end{equation}
where $z$ is the redshift  : $1+z=1/a(\tau)$, obtained by setting the current scale factor to unity and $\Omega$ is current observed value ($z=0$) of $\Omega(a)$. This completes the necessary review on the homogeneous cosmology front. It is clear that the $\Lambda{\rm CDM}$ limit corresponds to $\Omega_l \to 0$ or $l \to \infty$. Note also that in this limit we have $\beta \to -\infty$.

We are chiefly  interested in the theory of spherical, scalar perturbations predicted by this model, pertaining the large scale cosmic structures. So we next take in the  ansatz for the linear McVittie metric in \ref{bw2},
\begin{equation}
ds^2=a^2(\tau)\left[-(1+2\Phi (R,\tau))d\tau^2+(1-2\Psi(R,\tau))\left(dx^2+dy^2+dz^2 \right)\right]
\label{bw9'}
\end{equation}
where $R^2=x^2+y^2+z^2$ and $\Phi$ and $\Psi$ are the gravitational potentials. In the absence of anisotropic spatial stresses, we have $\Psi=\Phi$, which would not be the case here.  The  most general sources that can generate such spatial inhomogeneity are given by 
\begin{equation}
\delta {T^\mu}_\nu=\begin{bmatrix}
    -\delta\rho & -\rho\nabla_i u  \\
    \frac{\rho\nabla^i u}{a^2} & \delta P\, \delta^i{}_j +\frac{\zeta^i{}_j}{a^2}
\end{bmatrix}
\label{bw10}
\end{equation}
where $\zeta_{ij}=(\nabla_i \nabla_j- \delta_{ij}\nabla^2/3)\zeta$ ($i,~j \equiv x,y,z$) with $\zeta$ being a scalar used to parametrize the anisotropic strength tensor $\zeta_{ij}$. $u(R,\tau)$ is the velocity potential function ignoring any vorticity, $\delta \rho (R,\tau)$ is the perturbation representing the central overdensity and $ \delta P$ is the pressure perturbation.  We also have for the Weyl fluid perturbation,
\begin{equation}
m^2 \delta {\mathcal{C}^\mu}_\nu= \begin{bmatrix}
    -\delta\rho_\mathcal{C} & -(\rho_r+P_r)\nabla_i u_\mathcal{C}  \\
    \frac{(\rho_r+P_r)\nabla^i u_\mathcal{C}}{a^2} & \frac{\delta \rho_C \delta^i{}_j}{3} +\frac{\delta\pi^i{}_j}{a^2}
\end{bmatrix}
\label{bw11}
\end{equation}
where  $\rho_r$ and $P_r$ are the homogeneous density and pressure of radiation, $\delta_C=\frac{3\delta\rho_C}{4\rho_r}$  and $u_C$ is the velocity potential for the Weyl Fluid, also $ \delta \pi_{ij}=\left( \nabla_i \nabla_j -{\delta_{ij}}\nabla^2/3  \right)\delta \pi_{\mathcal{C}}$ and $ \delta \pi_\mathcal{C}$ is again a scalar used to parametrize $\delta \pi_{ij} $. We have
\begin{equation}
\frac{1}{a}  {{\dot\delta} _C}=\nabla^2 u_C 
\label{bw12}
\end{equation}
from which one gets
\begin{equation}
\frac{\ddot{\delta_C}}{a^2}+\left(\frac{2\beta}{\beta+2}-3\gamma\right)\frac{{\cal H}}{a^2} \dot{\delta_C}-\frac{1}{3}\left(2+3\gamma\right)\nabla^2 \delta_C
= \frac{1+3\gamma}{4\rho_r}\nabla^2 \left( \rho \Delta \right)
\label{bw13}
\end{equation}
where
\begin{equation}
\Delta=\delta+\frac{ 3{\mathcal H} u}{a}
\label{bw13'}
\end{equation}
with $\delta=\delta \rho/\rho$. We shall not require all the perturbation sources described above. In particular, since we are interested in the length scale of a structure which is essentially  subhorizon, we can safely ignore the temporal derivatives of the inhomogeneous perturbations compared to their spatial derivatives. 

Finally, we have the differential equations determining   the two potentials
\begin{equation}
\nabla^2\Psi = \frac{2+\beta}{2m^2 \beta} \rho\Delta + \frac{4\rho_r}{3m^2 \beta} \left(\delta_C + \frac{3{\mathcal H} u_C}{a}\right)
\label{bw14}
\end{equation}
and 
\begin{equation}
\nabla^2(\Psi-\Phi) = \frac{8\rho_r}{3m^2 \beta}\left[\delta_C+\frac{6 {\mathcal{H}}u_C}{a(2+\beta)}+\frac{3 \rho \Delta}{4\rho_r} \right]
\label{bw15}
\end{equation}
In the limit $\beta \to -\infty$, the right hand side of the above equation vanishes and we recover $\Lambda{\rm CDM} $, $\Psi=\Phi $. 

With all this equipment, we are now ready to go into the maximum turn around calculations.

\section{Calculation of the maximum turnaround radius}\label{sec.3}
We shall first demonstrate  the calculation of  the maximum turn around radius for a spherical cosmic structure ignoring the Weyl term to demonstrate the method. Let us consider a shell of backreactionaless cold dark matter fluid moving just outside the structure. Our starting point will be (see~\cite{PavlidouTetradisTomaras2014}) to consider the proper or physical  spatial coordinate corresponding to the cold dark matter's perturbation, $\delta \rho (R,\tau)$,
\begin{equation}
\vec{r}=a(\tau)\vec{x}
\end{equation}
The velocity and acceleration of this element with respect to the proper or physical time $dt=a(\tau) d\tau$ (since the cold dark matter is essentially non-relativistic) reads
\begin{equation}
\frac{dr^i}{dt}=\frac{1}{a(\tau)}\frac{d \vec{r}}{d \tau }=\delta u^i+{\mathcal H}x^i
\label{bwn1}
\end{equation}
where $\delta u^i $ is the peculiar  velocity. Using \ref{bw9'} we can derive the conservation equation for the perturbation
$$\delta {\dot {\vec {u}}}+\mathcal{H}\delta\vec{u}=-\vec{\nabla}\Phi $$
Differentiating \ref{bwn1}  with respect to the proper time once again and using the above equation, we obtain the acceleration
\begin{equation} 
\frac{d^2 \vec{r}}{dt^2}=\left(\frac{\ddot{a}}{a^3}-\frac{\dot{a}^2}{a^4}\right)\vec{r}-\frac{1}{a}\vec{\nabla}\Phi \equiv \frac{\dot{\mathcal{H}}}{a^2} \vec{r}-\frac{1}{a} \vec{\nabla} \Phi
\label{bw17}
\end{equation}
Note that the above relation is model independent. The explicit model dependence will enter via $\mathcal{H}$ and $\Phi$. 
Since length scales pertaining the structures are essentially sub-Hubble, the velocity potential for matter can usually be ignored and spatial derivatives of the perturbations will be favoured over time derivatives. Subtracting \ref{bw15} from \ref{bw14} and dropping all the Weyl terms give the Poisson equation for the potential $\Phi$,
\begin{equation}
\nabla^2 \Phi =\frac{2+\beta}{2m^2 \beta}\delta \rho - \frac{2}{m^2\beta}\delta \rho
\label{bw17.5}
\end{equation}
where we have ignored  the velocity perturbation, i.e. $ \rho\Delta \sim \delta\rho$ in \ref{bw13'}.  Thus we get
\begin{equation}
\nabla^2 \Phi = 4\pi G_{\rm eff} \delta\rho
\label{bw18}
\end{equation}
where 
\begin{equation}
G_{\rm eff}=G \left( 1-\frac{2}{\beta}\right) =G\left(1+\frac{1}{1+\frac{l \mathcal{H}}{a}}\right) \qquad {\rm where } \qquad G \equiv \frac{1}{8\pi m^2}
\label{bw19}
\end{equation}
where we have also used \ref{bw8}.
$G_{\rm eff}$ approaches $G$ as $l \to \infty$, the $\Lambda{\rm CDM}$ limit. Thus in this theory the effective Newton's `constant' is larger than $G$, indicating the increase of gravitational attraction. We next approximate the whole structure  as a point  mass located at $\vec{r}=0$ : $\delta \rho =M\delta^3(\vec{r})$ acts as the perturbation,
\begin{equation}
\nabla^2 \Phi= 4\pi G_{\rm eff} M \delta^3 (\vec{R}a(\tau))
\label{bw20}
\end{equation}
giving
\begin{equation}
\Phi=-\frac{G_{\rm eff}M}{R}
\label{bw21}
\end{equation}
The maximum turn around radius  $R_{\rm TA, max}$ is by definition the point of vanishing acceleration. Thus setting $d^2 \vec{r}/dt^2=0$ in \ref{bw17} and noting that from the spherical symmetry of the problem we have $\vec{r}\equiv r$, we obtain
\begin{equation}
\frac{\mathcal{\dot{H}}\, R_{\rm TA, max}}{a}-\frac{G_{\rm eff}M}{R_{\rm TA, max}^2}=0
\end{equation}
Using now \ref{fried} along with the homogeneous conservation equation,  $\dot{\rho}+3\mathcal{H}\rho=0 $, and
\begin{equation}
\dot{\mathcal{H}}=\mathcal{H}^2+\frac{1}{6m^2}\frac{al\dot{\rho}}{\sqrt{1+l^2\left(\frac{\rho+\sigma}{3m^2} \right)}}
\label{bwhdot}
\end{equation}
we finally arrive at
\begin{equation}
R_{\rm TA,max}=\left(\frac{G_{\rm eff}M}{\frac{1}{l^2}\left(-1+\sqrt{1+l^2\frac{\rho+\sigma}{3m^2}}\right)^2-\frac{\rho}{2m^2}\left(1-\frac{1}{\sqrt{1+l^2\frac{\rho+\sigma}{3m^2}}}\right)}\right)^{1/3}
\label{bw22}
\end{equation}
We recover the $\Lambda{\rm CDM}$ result by setting  $l \to \infty$ above,
\begin{equation}
R_{\rm TA,max}=\left(\frac{GM}{\frac{\Lambda_\sigma}{3}-\frac{\rho}{6m^2}}\right)^{1/3}
\label{bw23}
\end{equation}
where we have written $\Lambda_{\sigma}=\sigma/m^2$ for the brane cosmological constant. We can rewrite the above equation as
\begin{equation}
R_{\rm TA,max}=\left(\frac{3GM}{\Lambda_\sigma}\right)^{1/3}\left(1-\frac{\rho}{2m^2\Lambda_\sigma}\right)^{-1/3}
\label{bw23'}
\end{equation}
We may just include the background density $\rho$ in the mass term via the redefinition  $M^{\prime} =M\left(1+\rho/2m^2\Lambda_\sigma \right) $, and arrive at
\begin{equation}
R_{\rm TA,max}=\left(\frac{3GM^{\prime}}{\Lambda_\sigma}\right)^{1/3}
\label{bw24}
\end{equation}
The mass function $M'$ clearly  should be regarded as a total or effective mass function, taking into account the effect of the homogeneous matter density as well. For nearby cosmic structures, which is our main focus, we may take $z\sim 0$. Then recalling $1/2m^2\equiv 4\pi G$, using $\Omega_{m} \simeq 0.3$ and $\Omega_{\Lambda_\sigma} \simeq 0.7$ for ${\Lambda{\rm CDM}}$, it is easy to see that $M' \approx 1.214M$. One then uses the observed  mass versus actual size data to do phenomenology in this context~\cite{PavlidouTetradisTomaras2014}, which we shall be explicitly discussing at the end of Sec.~3.1.\\

\noindent
Next we shall include the effect of the inhomogeneous  Weyl fluid to investigate how it modifies \ref{bw22}, while keeping still $\Lambda=0$.
This would  simply correspond to  modifying $G_{\rm eff}$ as follows. First, we recall that we already have ignored the homogeneous cosmological part of it (\ref{sec.2}). The velocity potential of the perturbation of the fluid satisfies
$\nabla^2 u_C=0$, as we may ignore the temporal variation with respect to the spatial ones  in \ref{bw12}, in the subhorizon length scale we are concerned with. Next note that  in \ref{bw14} and \ref{bw15}, $u_C$ comes multiplied with the homogeneous radiation density, $\rho_r$, which has little effect in the self gravity effects of a large scale structure. Also from \ref{bw13}, we have
\begin{equation}
\delta \rho_C = - \frac{1+3\gamma}{2+3\gamma} \delta \rho
\label{bw25}
\end{equation}
where we have set an additive integration constant to zero, as the inhomogeneity is by definition sourced by the central overdensity. Subtracting now \ref{bw15} from \ref{bw14},  we obtain 
\begin{equation}
\nabla^2 \Phi = 4\pi G\left(1- \frac{2}{\beta\left(3\gamma +2 \right)}\right)\delta\rho
\label{bw26}
\end{equation}
We have from \ref{bw9} $\beta\leq 0$ and
\begin{equation}
3\gamma+2 =3 -\frac{3\Omega_m(1+z)^3}{2\left[ \Omega_m(1+z)^3+\Omega_{\sigma}+\Omega_l\right]}>0,~~{\rm always}
\label{bw26'}
\end{equation}
In other words, the effective Newton's `constant'
\begin{equation}
G_{\rm eff}=G\left(1- \frac{2}{\beta\left(3\gamma +2 \right)}\right)
\label{bw27}
\end{equation}
appearing in \ref{bw26} is always larger than that of $\Lambda{\rm CDM}$, as earlier. Also, $G_{\rm eff}$ reduces to  $G$ in the limit $l \to \infty$ (or $\beta \to -\infty$). 

Let us now compute $R_{\rm TA,max}$ for this case and compare the result with  $\Lambda{\rm CDM}$. We rewrite  \ref{fried} as
\begin{equation}
\left(\frac{\mathcal{H}}{a}\right)^2=\frac{\rho+\sigma}{3m^2}+\frac{2}{l^2}\left[1-\sqrt{1+l^2\left(\frac{\rho+\sigma}{3m^2}\right)}\right]
\label{bw28}
\end{equation}
Using \ref{bw8} and the density parameter corresponding to $l$, we get
\begin{equation}
\left(\frac{\mathcal{H}}{a}\right)^2=\frac{\rho+\sigma}{3m^2}-\frac{2\mathcal{H}}{l a}=\frac{\rho+\sigma}{3m^2}\frac{1}{1+2\sqrt{\Omega_l}}
\label{bw30}
\end{equation}
We also note that 
\begin{equation}
1-\frac{1}{\sqrt{1+l^2(\frac{\rho+\sigma}{3m^2})}}=1-\frac{1}{1+\frac{l \mathcal{H}}{a}}=1-\frac{\sqrt{\Omega_l}}{\sqrt{\Omega_l}+1}
\label{bw31}
\end{equation}
Since $\sqrt{\Omega_l}$ is expectedly a `small' number, we shall now proceed perturbatively in it.
Using the above  equations and \ref{bw17}, \ref{bwhdot},  we obtain after a lengthy but straightforward computations, up to  the leading order in $\Omega_l$,
\begin{equation}
R_{\rm TA,max}=\left[\frac{G_{\rm eff}M}{\frac{\sigma}{3m^2}-\frac{\rho}{6m^2}-\frac{2\sigma\sqrt{\Omega_l}}{3m^2}-\frac{\rho \sqrt{\Omega_l}}{2m^2}}\right]^{1/3}
\label{bw32}
\end{equation}
Recalling $\Lambda_{\sigma}=\sigma/m^2$, we now compare the above expression with \ref{bw23} corresponding to $\Lambda {\rm CDM}$.  We have already proven that $G_{\rm eff} \geq G$ and the denominator of the above equation is obviously smaller than that of \ref{bw23}. Since $\sqrt{\Omega_l}$ is a small number and $\beta \sim \Omega_l^{-1/2}$, \ref{bw9}, we can  further express the leading corrections to $R_{\rm TA, max}$ as
\begin{equation}
R_{\rm TA,max}\approx  \left(\frac{3GM'}{\Lambda_{\sigma}} \right)^{1/3} \left[1+\frac{2\sqrt{\Omega_l}}{3}+\frac{\sqrt{\Omega_l} \rho}{2m^2\Lambda_{\sigma}}-\frac{2}{3\beta(3\gamma+2)}\right]
\label{bw32'}
\end{equation}
where $M'=M(1+\rho/2m^2\Lambda_{\sigma})$ as earlier. Thus by \ref{bw26'} and since $\beta<0$,  the maximum turn around radius  or the maximum possible size of a cosmic structure predicted by this theory is larger than $\Lambda{\rm CDM}$, provided that $\Lambda_{\sigma}$ equals (or is smaller than) the observed value of the cosmological constant in $\Lambda{\rm CDM}$. However, note that in order to be in the complete braneworld scenario, we must treat its parameter space independently and should permit  values of $\Lambda_{\sigma}$ (as well $\rho$) different from $\Lambda{\rm CDM}$, both  larger and smaller. We shall perform this more general analysis numerically and {\it  non-perturbatively} in the next section and will see that  as long as the bulk $\Lambda$ is vanishing, the above conclusion remains the same. 
Moreover, we shall `turn on' a bulk $\Lambda$ below as well and owing to the complicated structure of the field equations, we will be able to obtain clear constraints to rule out certain regions of the parameter space of the theory, based upon the current mass versus size data.

\subsection{Inclusion of a bulk cosmological constant}\label{sec.3.1}
We shall now be needing \ref{bw7} along with the generalized form of \ref{bw8}, 
\begin{equation}
\beta=-2\sqrt{1+l^2\left(\frac{\rho+\sigma}{3m^2}-\frac{\Lambda}{6}\right)} \qquad 
3\gamma-1=-\frac{\rho}{2m^2\left(\frac{\rho+\sigma}{3m^2}+\frac{1}{l^2}-\frac{\Lambda}{6}\right)}
\label{bw40}
\end{equation}
The generalization to the effective Newton's constant comes readily substituting $\beta$ and $\gamma$ from \ref{bw40} into \ref{bw26} or \ref{bw27} and  our $R_{\rm TA, max}$ will be given by
\begin{equation}
R_{\rm TA,max}=\left(\frac{G_{\rm eff}M}{{\mathcal{\dot{H}}}/{a^2}}\right)^{1/3}
\label{bw41}
\end{equation}  
Following similar steps as earlier, we now obtain a modified expression for $R_{\rm TA, max}$, incorporating the effect off the bulk cosmological constant, 
\begin{equation}
R_{\rm TA,max} = \left(\frac{3GM}{\Lambda_\sigma} \frac{1- \frac{2}{\beta\left(3\gamma +2 \right)}}{1-\frac{\rho}{2m^2\Lambda_{\sigma}}+\frac{6}{l^2  \Lambda_\sigma}\left[1-\sqrt{1+l^2\left(\frac{\rho+m^2 \Lambda_{\sigma}}{3m^2}-\frac{\Lambda}{6}\right)}\right]+\frac{3\rho}{2m^2\Lambda_\sigma}\frac{1}{\sqrt{1+l^2\left(\frac{\rho+m^2\Lambda_{\sigma}}{3m^2}-\frac{\Lambda}{6}\right)}}}\right)^{1/3}
\label{bw42}
\end{equation}
Expressing  this with respect to the primed mass $M^\prime=M\left(1+\rho/2m^2\Lambda_\sigma \right) $, we  obtain 
\begin{equation}
R_{\rm TA,max} = \left(\frac{3GM^\prime}{\Lambda_\sigma} \frac{1- \frac{2}{\beta\left(3\gamma +2 \right)}}{1-\frac{\rho}{2m^2\Lambda_{\sigma}}+\frac{6}{l^2  \Lambda_\sigma}\left[1-\sqrt{1+l^2\left(\frac{\rho+m^2\Lambda_{\sigma}}{3m^2}-\frac{\Lambda}{6}\right)}\right]+\frac{3\rho}{2m^2\Lambda_\sigma}\frac{1}{\sqrt{1+l^2\left(\frac{\rho+m^2\Lambda_{\sigma}}{3m^2}-\frac{\Lambda}{6}\right)}}}\right)^{1/3}\left(\frac{1}{1+\rho/2m^2\Lambda_\sigma} \right)^{1/3}
\label{bw42.5}
\end{equation}
From \ref{bw7} and using  $\Omega_l=\frac{a^2}{l^2 \mathcal{H}^2}$, we have
\begin{equation}
\sqrt{1+l^2\left(\frac{\rho+\sigma}{3m^2}-\frac{\Lambda}{6}\right)}=\sqrt{\frac{1}{\Omega_l}-\frac{l^2 \Lambda}{6}}+1
\label{bw43}
\end{equation}
Let us define   
\begin{equation}
R_{\Lambda{\rm CDM}}:=\left(\frac{3GM^\prime}{\Lambda_0}\right)^{1/3}
\label{new1}
\end{equation}
where $\Lambda_0$ denotes the value of the cosmological constant fitted to $\Lambda{\rm CDM}$ $\sim 10^{-52}{\rm m}^{-2}$. Thus the above is nothing but the $R_{\rm TA, max}$ of $\Lambda{\rm CDM}$. Using this and \ref{bw43}, we reexpress \ref{bw42.5} as  
\begin{equation}
\frac{R_{\rm TA,max}}{R_{\Lambda{\rm CDM}} }=\left(\frac{1- \frac{2}{\beta\left(3\gamma +2 \right)}}{1-\frac{\rho}{2m^2 \Lambda_{\sigma}}-\frac{6\Omega_l \mathcal{H}^2}{a^2\Lambda_\sigma}\sqrt{\frac{1}{\Omega_l}-\frac{\Lambda a^2}{6\Omega_l \mathcal{H}^2}}+\frac{3\rho}{2m^2\Lambda_\sigma}\frac{1}{\sqrt{\frac{1}{\Omega_l}-\frac{a^2\Lambda}{6\Omega_l \mathcal{H}^2}}+1}}\right)^{1/3}\left(\frac{\Lambda_0}{\Lambda_{\sigma}(1+\rho/2m^2\Lambda_\sigma)} \right)^{1/3} 
\label{new3}
\end{equation}
Thus the $R_{\rm TA,max}$ of $\Lambda{\rm CDM}$ serves as a point of reference above and we shall investigate how much the above ratio could deviate from unity, subject to different values of all parameters of the model ($\rho,\,\Lambda_{\sigma},~\Lambda,~\Omega_l$), {\it independent of} ${\Lambda{\rm CDM}}$. 

First we note that 
the ratio $\rho/m^2\Lambda_{\sigma}$ above could be replaced with $\Omega_m/\Omega_{\sigma}$. 
\begin{figure}[h]
\includegraphics[width=1.3\textwidth, center]{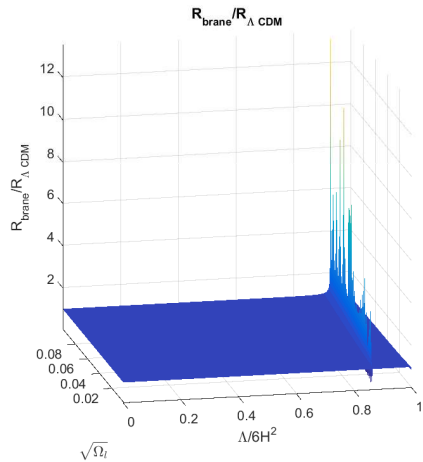}
\caption{A 3-dimensional plot of \ref{new3} with respect to the variation of the (dimensionless) parameters with a {\it positive} bulk cosmological constant. The current scale factor is set to unity and we have taken $ \Omega_m\simeq0.27$ as the only input. $\Omega_{\sigma}$ or $\Lambda_{\sigma}$ at each point is given via \ref{bw45}. Chiefly, a) for $\Lambda=0$ the ratio is always larger than $\Lambda{\rm CDM}$, proving our earlier claim made at the end of \ref{sec.3} and b) parameter regions where the ratio could be smaller are indicated and more clearly demonstrated in \ref{f2} and subsequent plots with $\Omega_m$ values different from $\Lambda{\rm CDM}$. }
\label{f1}
\end{figure}
\begin{figure}[h]
\includegraphics[width=1\textwidth, center]{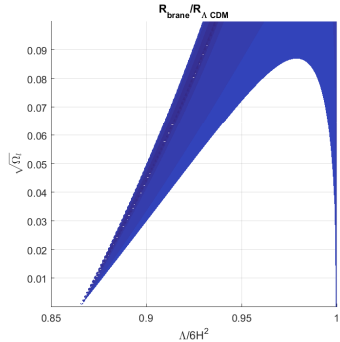}
\caption{Plot of \ref{new3} to demonstrate  the region of parameters excluded by mass versus actual size observations (the coloured region).   }
\label{f2}
\end{figure}
We now wish to make plots of \ref{new3} subject to the variation of the bulk cosmological constant and the parameter $\Omega_l$, by taking various values of
$\Omega_m$ as the input.  Let us divide both sides of \ref{bw7} with $({\mathcal{H}}/{a})^2$ and use \ref{bw43} to get
\begin{equation}
1-\Omega_m = \Omega_\sigma-2 \Omega_l \sqrt{\frac{1}{\Omega_l}-\frac{a^2\Lambda}{6\Omega_l \mathcal{H}^2}} 
\label{bw45}
\end{equation}
We shall use different values of $\Omega_m$ in the above equation and using that would  replace $\Omega_{\sigma}$ (or $\Lambda_{\sigma}$) in \ref{new3} in the favour of $\Omega_l$.  

\begin{figure}[h]
\makebox[\textwidth]{%
\begin{adjustbox}{center,max width=0.6\textwidth}
\includegraphics[width=1.5\textwidth]{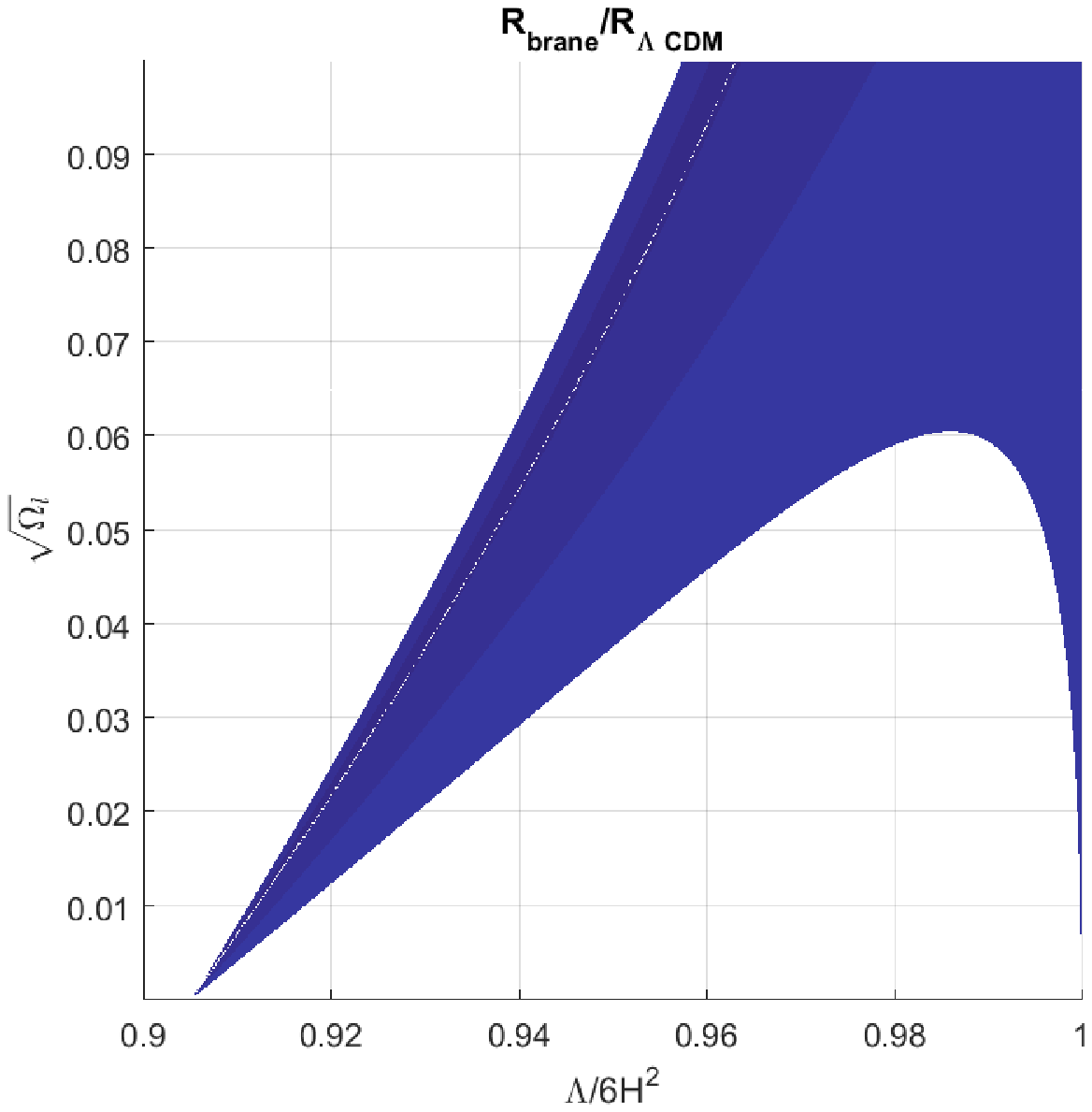}%
\end{adjustbox}
\begin{adjustbox}{center,max width=0.6\textwidth}    
\includegraphics[width=1.5\textwidth]{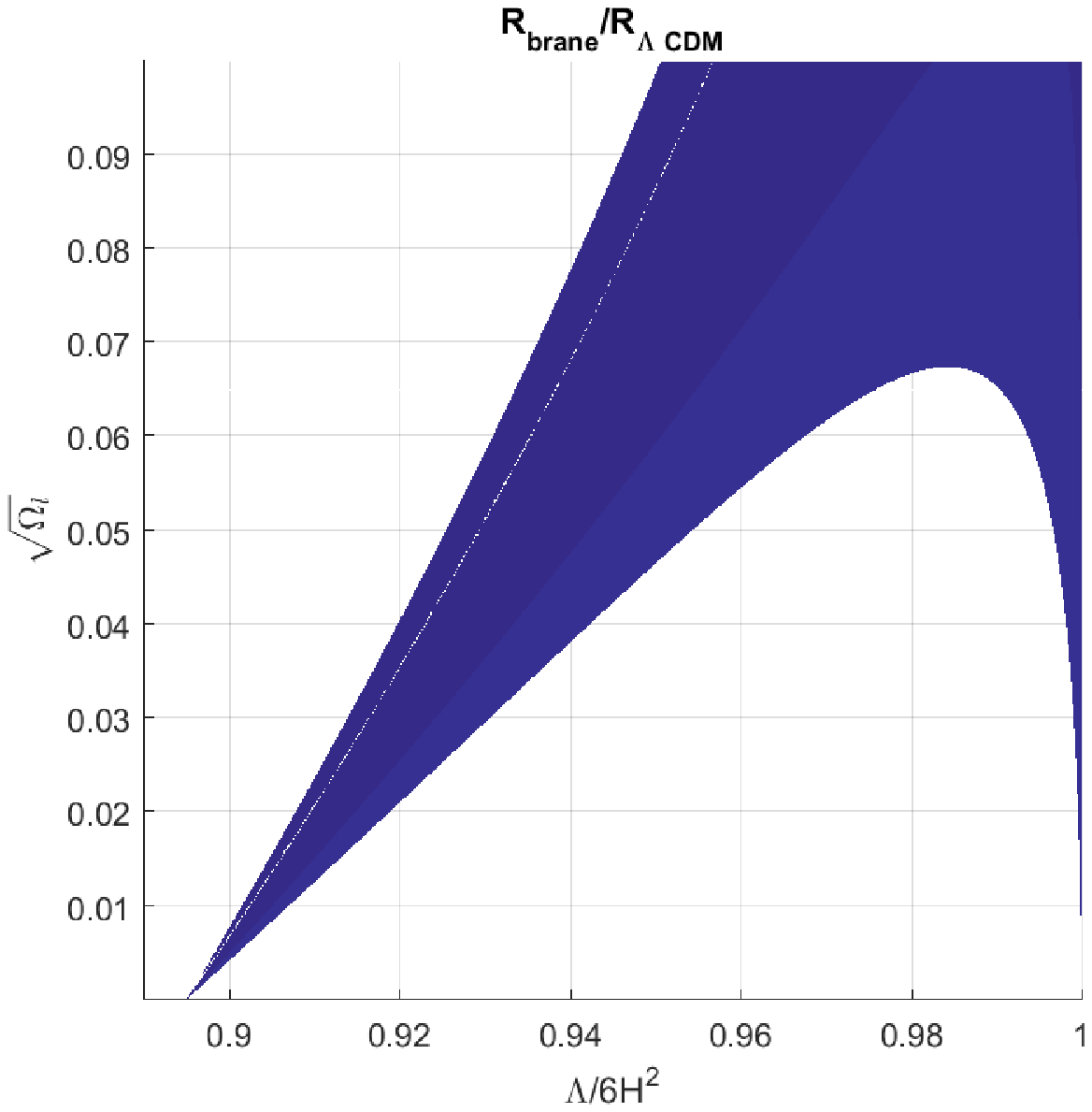}%
\end{adjustbox}
}\\[0.5cm]
\makebox[\textwidth]{%
\begin{adjustbox}{center,max width=0.6\textwidth}
\includegraphics[width=1.5\textwidth]{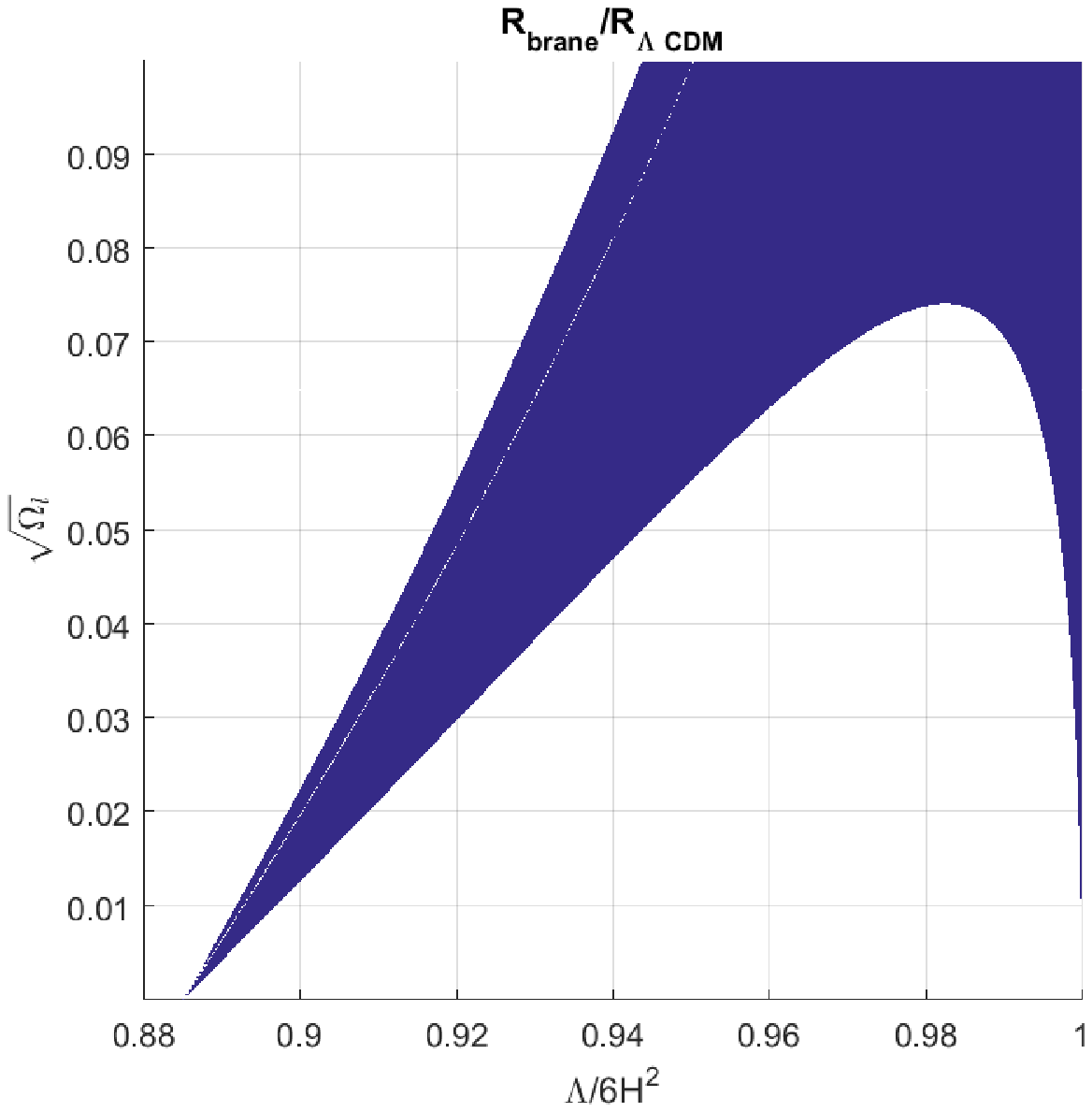}%
\end{adjustbox}
\begin{adjustbox}{center,max width=0.6\textwidth}    
\includegraphics[width=1.5\textwidth]{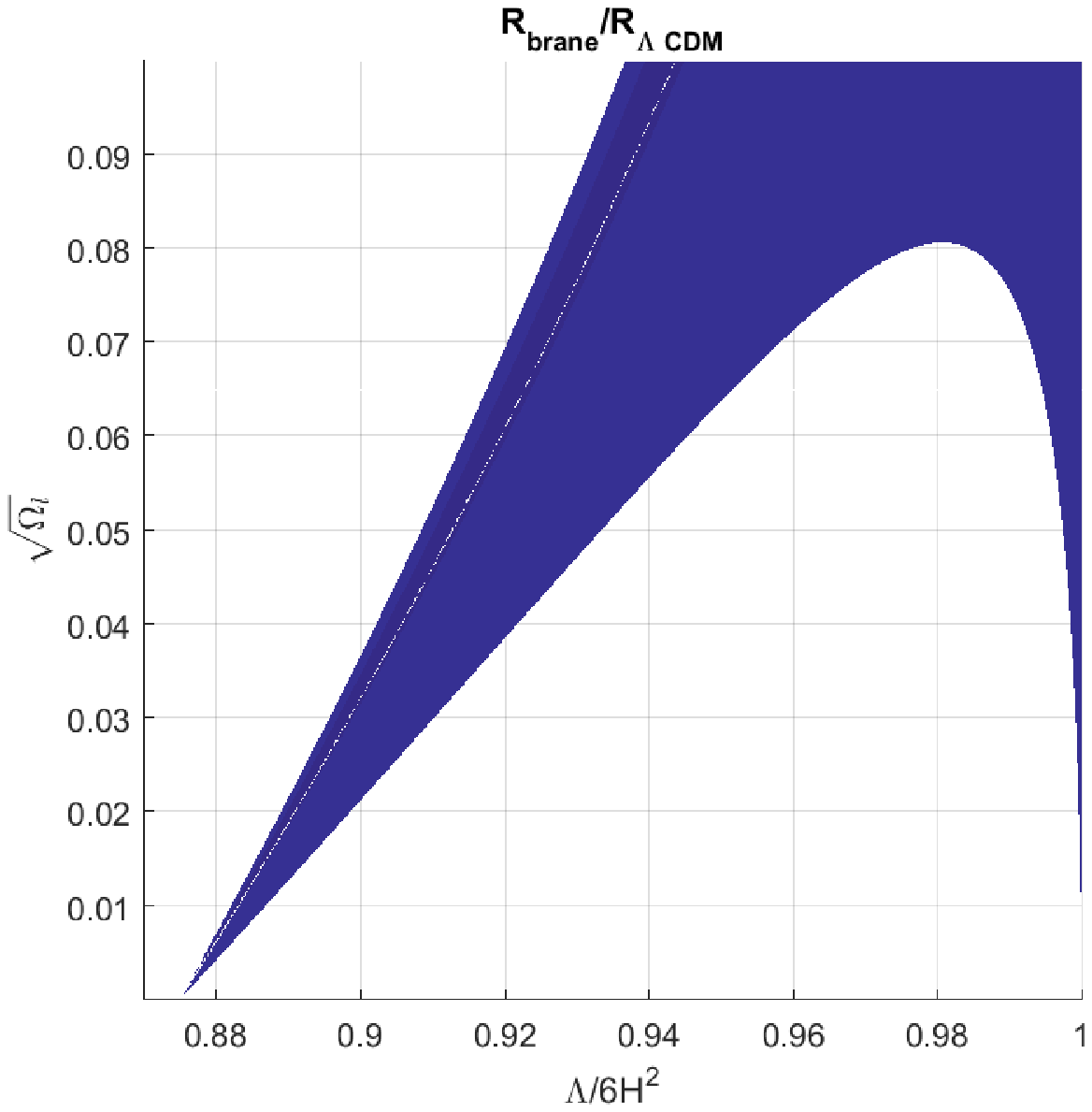}%
\end{adjustbox}
}\\[0.5cm]
\makebox[\textwidth]{%
\begin{adjustbox}{center,max width=0.6\textwidth}
\includegraphics[width=1.5\textwidth]{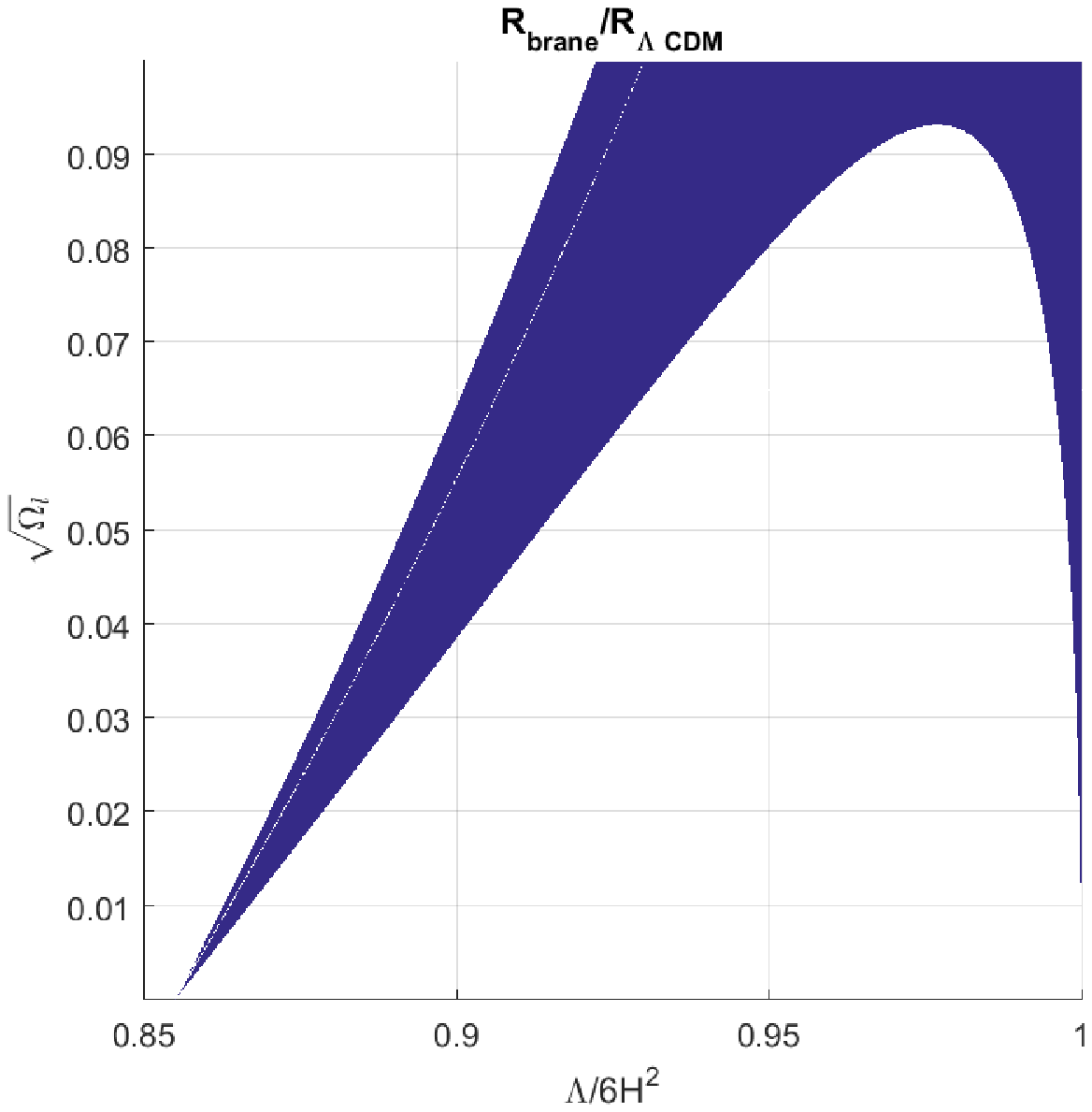}%
\end{adjustbox}
\begin{adjustbox}{center,max width=0.6\textwidth}    
\includegraphics[width=1.5\textwidth]{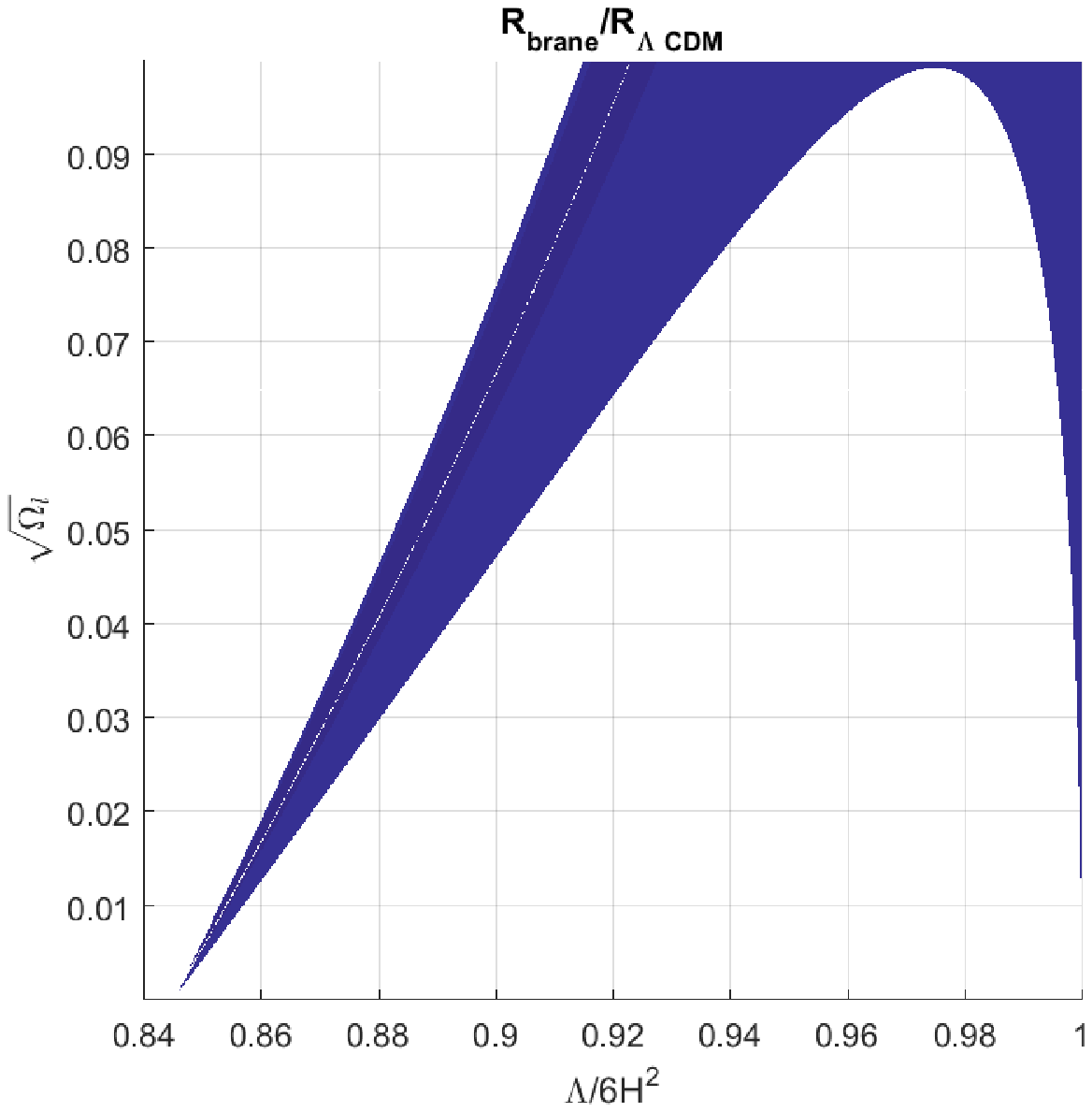}%
\end{adjustbox}
}\\[0.5cm]
\makebox[\textwidth]{%
\begin{adjustbox}{center,max width=0.6\textwidth}
\includegraphics[width=1.5\textwidth]{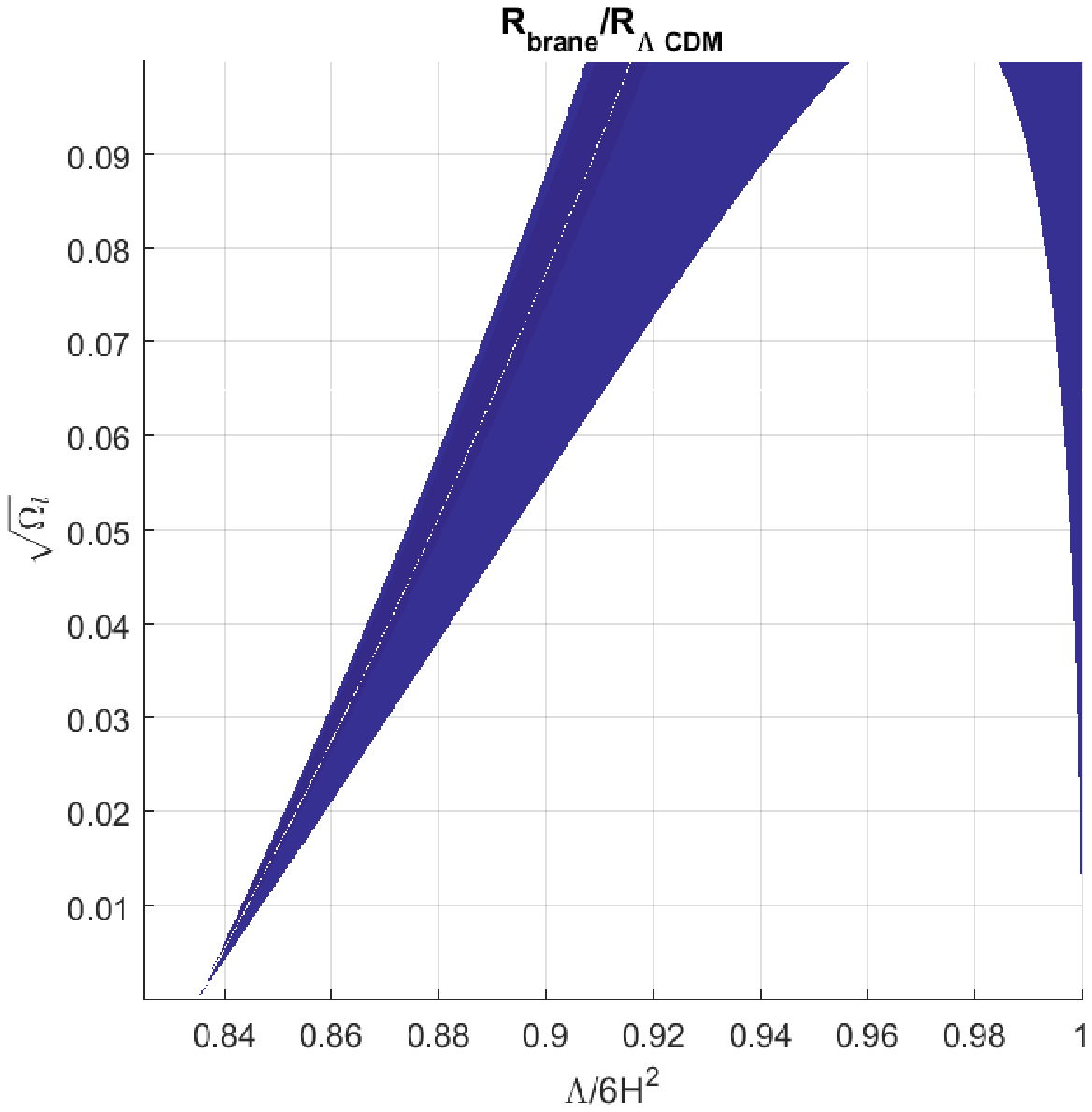}%
\end{adjustbox}
\begin{adjustbox}{center,max width=0.6\textwidth}    
\includegraphics[width=1.5\textwidth]{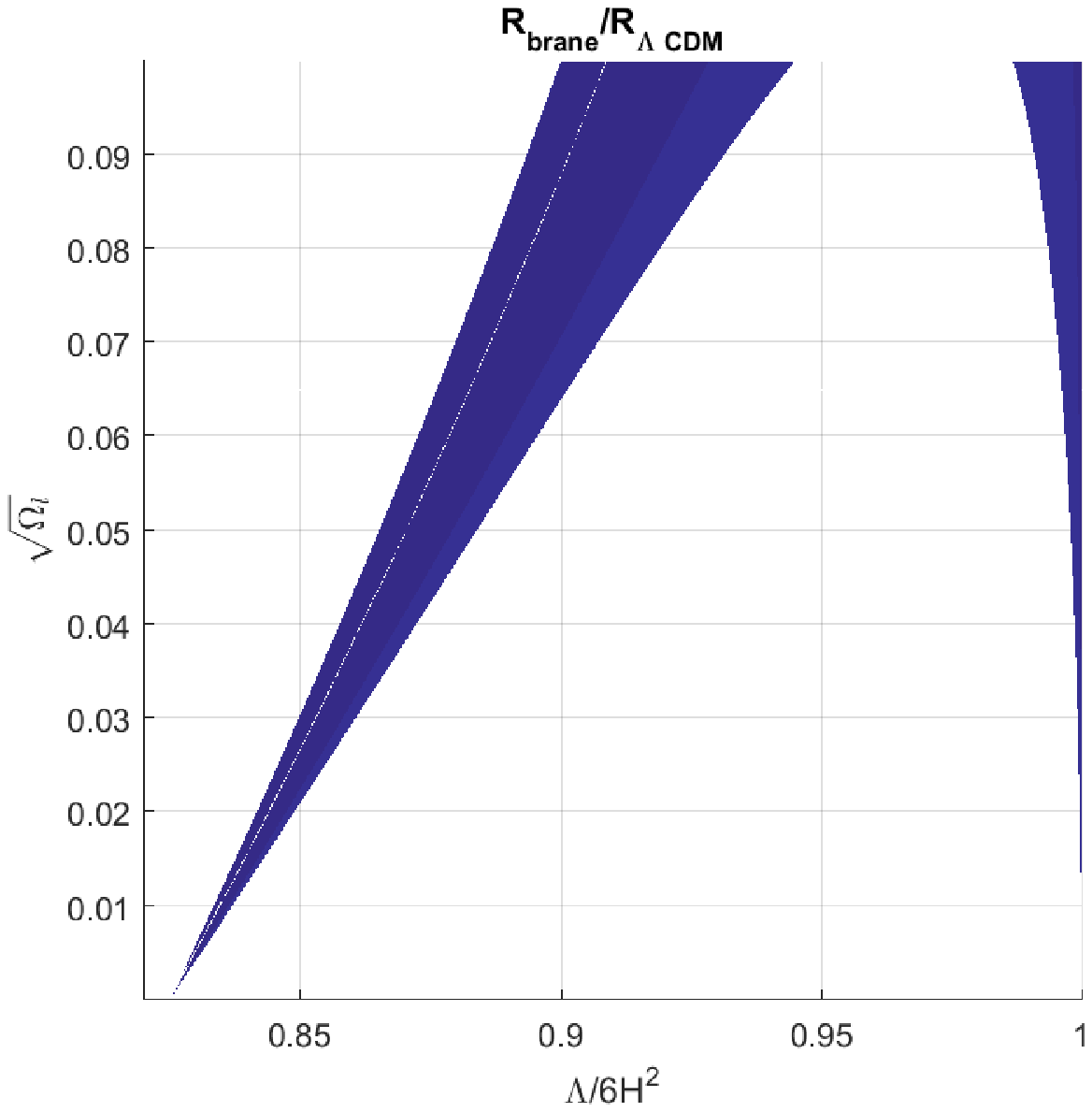}%
\end{adjustbox}
}\\[0.5cm]
\caption{Plots analogue to that of \ref{f2} but with $ \Omega_m=$ 0.19, 0.21, 0.23, 0.25, 0.29, 0.31, 0.33 and 0.35 respectively, moving from left to right. All these plots show as we increase $\Omega_m$, more region of the parameter space gets opened up for stable structures due to increased attractive effects of the cold dark matter. Note also that for $\Lambda=0$, there is no constraint.}
\label{f4}
\end{figure}

 To start with, let us take the $\Lambda{\rm CDM}$ value, $\Omega_m=0.27$.
\ref{f1} then depicts the variation of the ratio on the left hand side of \ref{new3}, with respect to the variation in the parameter space of the bulk $\Lambda$ and $\Omega_l$. As we have stressed earlier, for some {\it nearby, non-virial} cosmic structures with $M\gtrsim 10^{14}M_{\odot}$, the theoretical prediction of $\Lambda{\rm CDM}$ on $R_{\rm TA, max}$ is only roughly about $10\%$ larger than their actual observed sizes i.e. $ R_{\rm TA,obs}\simeq 1.1 R_{\Lambda{\rm CDM}}$~\cite{PavlidouTomaras2013, PavlidouTetradisTomaras2014}. Thus {\it any } alternative dark energy/gravity model predicting an $R_{\rm TA, max}$ lesser than about $10\%$ compared to  $\Lambda{\rm CDM}$ (i.e., $R_{\rm brane}< 0.9 R_{\Lambda{\rm CDM}}$), gets ruled out.  Based on that, we get the constraints on the parameter space of the theory depicted in \ref{f1} or more clearly in \ref{f2}. It is also evident that for $\Lambda=0$,
there is no constraint whatsoever, proving our earlier claim made at the end of \ref{sec.3}, regarding the consistency of this model with the bound of $R_{\rm TA, max}$ with a vanishing bulk $\Lambda$. Further in \ref{f4} we find analogues of \ref{f2} for values of $\Omega_m$ different from $\Lambda{\rm CDM}$.

\begin{figure}[h]
\includegraphics[width=.5\textwidth, center]{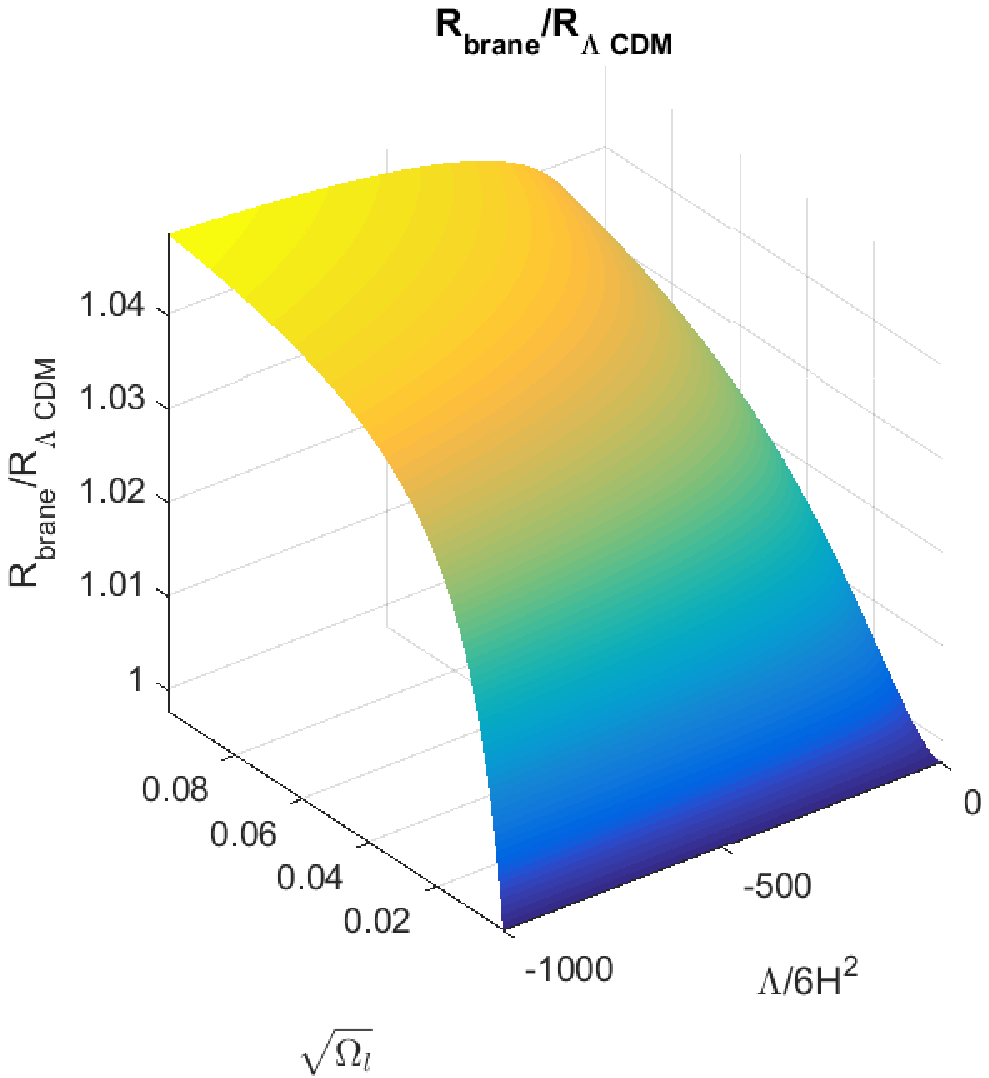}
%
\includegraphics[width=.5\textwidth, center]{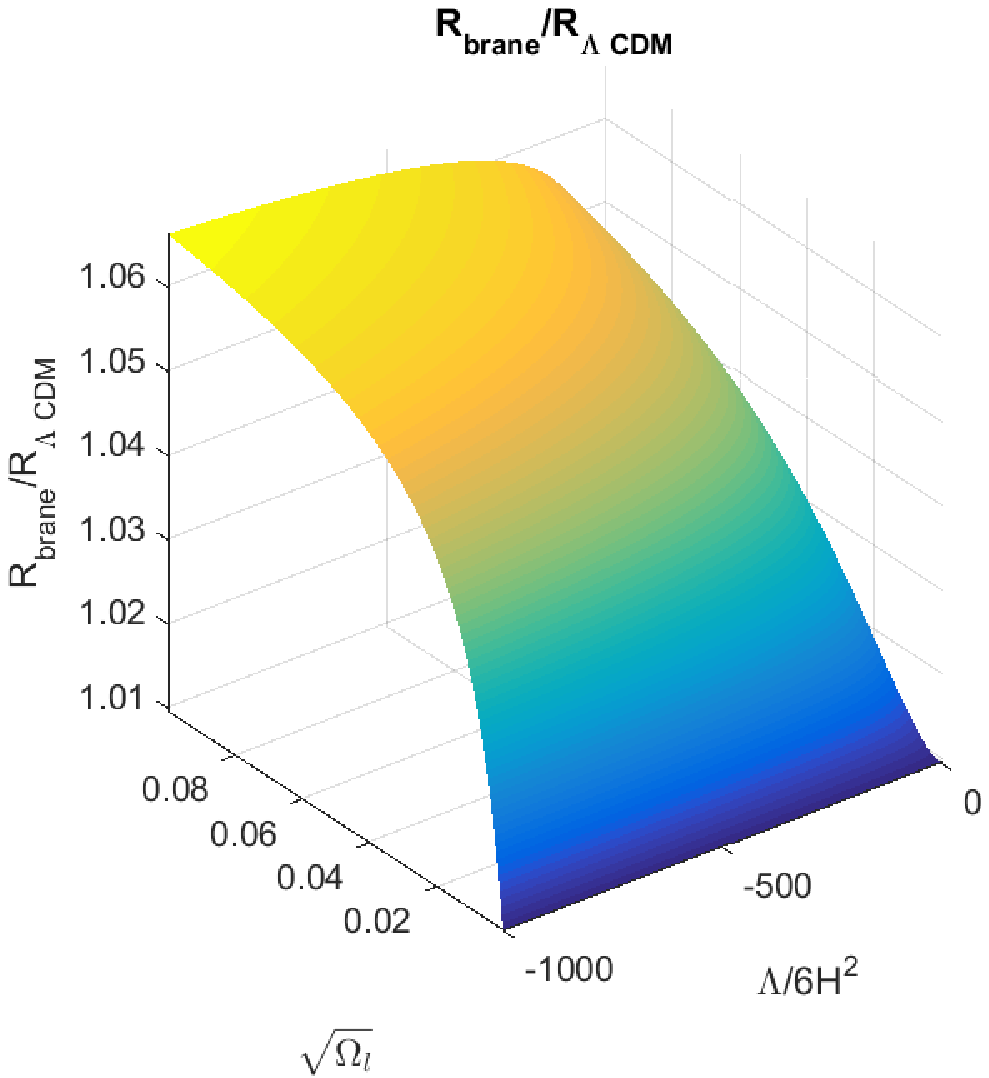}
%
\includegraphics[width=.5\textwidth, center]{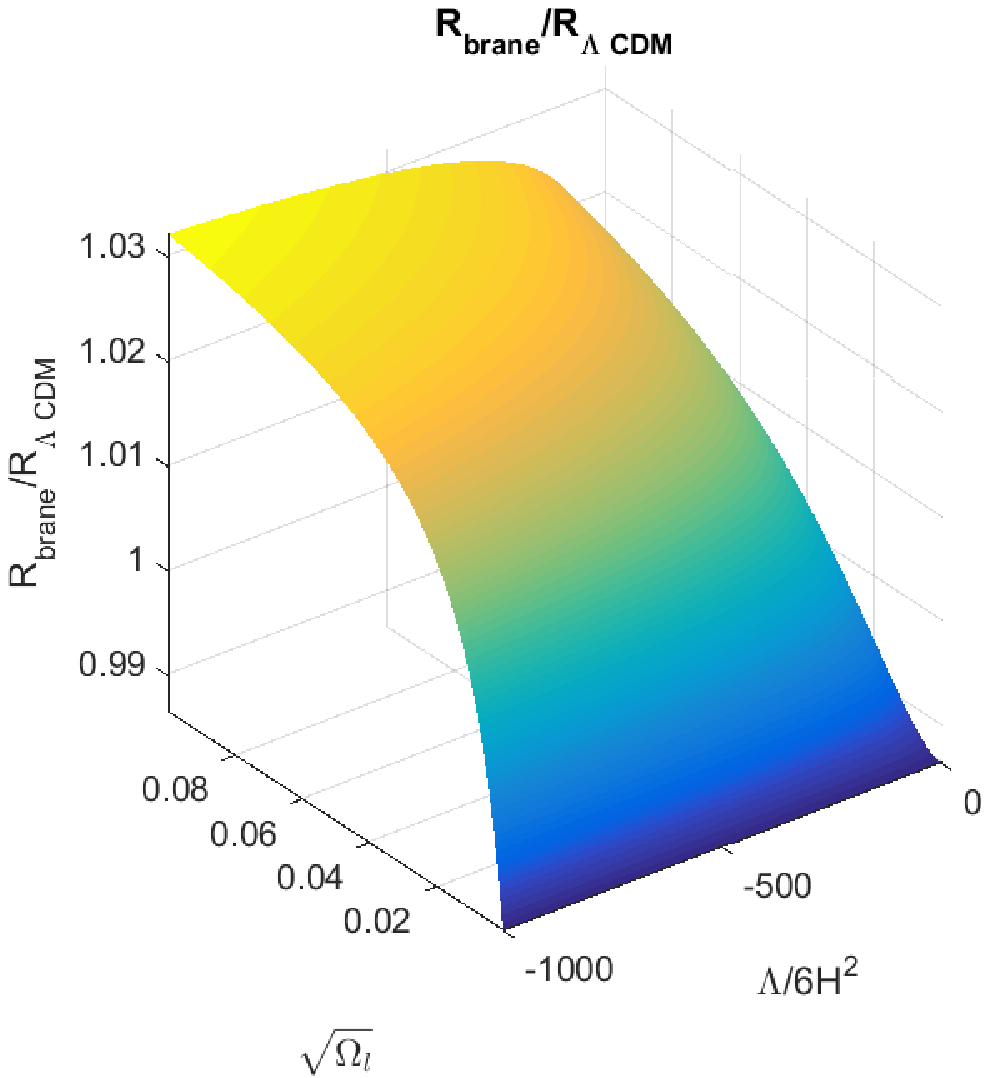}
\caption{Plots with a negative bulk $\Lambda$, respectively with  $ \Omega_m=0.27, 0.29, 0.25$, giving no constraints whatsoever.   We have checked that unless  $ \Omega_m $ is as small as $0.1$, the conclusion remains the same. The contrast between the results of positive and negative $\Lambda$ corresponds to the fact that the second produces attractive gravity as opposed to the first. Hence one should always expect prediction of bigger structure sizes for a negative bulk $\Lambda$, compared to $\Lambda \geq 0 $. However, note that the rather high magnitude of $\Lambda/6H^2$ above could be expected to be ruled out by some different cosmological observable(s), which unlike $R_{\rm TA, max}$ are sensitive to the absolute value of $\Lambda$ (see the end of \ref{sec.3.1} for discussion). Nevertheless, we have considered such values to demonstrate  the consistency with our  expectation pertaining the attractive gravitational effects created by a negative cosmological constant.}
\label{f6}
\end{figure}

Clearly, the situation should be the opposite if we flip the sign of the bulk $\Lambda$, as a negative $\Lambda$ creates attractive gravity as opposed to the positive one, thereby predicting larger structure sizes. 
Indeed, in \ref{f6}, we have numerically depicted this scenario with respect to independent parameters, with $\Lambda/6{ H}^2$ as high as up to $\sim -1000$.  We have checked that the conclusion remains the same for $\Omega_m$ as tiny as $0.1$.

By analysing the baryon acoustic oscillation and the cosmic microwave background data {\it without} any bulk $\Lambda$ for this model, it has been shown recently in~\cite{Alam:2016wpf} that $\Omega_l \lesssim 0.1 $.  However, if we set $\Lambda=0$ in our analysis, we have no constraint on $\Omega_l$, as we have seen that (\ref{f2}, \ref{f4}) in this case the predicted structure size is larger than what is actually observed. In fact if a certain theory predicts a structure size larger than what is actually observed, we cannot use $R_{\rm TA, max}$ to constrain it.   Such feature, in particular, is reflected into  \ref{f6}, where we have considered $|\Lambda/6H^2|$ as high as ${\cal O}(10^3)$ in order to demonstrate the consistency of $R_{\rm TA, max}$ with the fact that a negative $\Lambda$ creates attractive gravity. Note that this is unlike the case of a positive $\Lambda$, \ref{f2}, \ref{f4}, where such higher values are automatically ruled out by \ref{bw45}, in which the square root yields complex values of $\Omega_m$ for arbitrary large values of $\Lambda/6$, scaled by the Hubble rate squared. It then seems natural to expect that other cosmological/astrophysical observables which are sensitive to the absolute value of $\Lambda$ (unlike $R_{\rm TA, max}$) could rule out such higher values of a negative $\Lambda$. It could include usual observables like the luminosity and distance measures or the light bending angle at galactic scales and so on, e.g.~\cite{Weinberg:2008zzc}. It could also involve, finding out a static and spherically symmetric geometry corresponding to \ref{bw2} and to compute the perihelion precession. It seems very interesting and important to pursue such investigation in detail. 
%
\section{Summary and outlook}\label{sec.4}
In this work we have investigated the phantom braneworld model to compute the maximum turn around radius for a spherical cosmic structure with a given mass and to constrain it through the observed mass versus size data for them, for certain {\it nearby, non-virial} cosmic structures (see~\cite{PavlidouTomaras2013, PavlidouTetradisTomaras2014} and references therein). As we discussed in~\ref{sec.1}, it turns out that if a given theory predicts the size of such a structure to be smaller than around $10\%$ of $\Lambda{\rm CDM}$, it gets severely constrained on the basis of the stability.
We have shown that  $R_{\rm TA,max}$  found in \ref{new3} can actually go considerably smaller for a positive bulk $\Lambda$, in some region of the parameter space,  than the observed sizes, leading to interesting constraints on the parameter space of the theory, \ref{f1}, through \ref{f4}. We also have seen that for a negative or vanishing bulk $\Lambda$, we obtain no constraint, whatsoever. 

 We have used set of different values of $\Omega_m$, both larger and smaller than that of  $\Lambda{\rm CDM}$, $\Omega_m\simeq 0.27$, in \ref{new3}, in order to remain in the complete braneworld scenario. Likewise, for each such $\Omega_m$  value, $\Lambda_{\sigma}$ or $\Omega_{\sigma}$ is computed numerically from \ref{bw45}. From the plots with a positive bulk $\Lambda$, we have a clear pattern that as $\Omega_m$ increases, more and more region of the parameter space is opened up for a stable structure. This should be the manifestation of the increasing attractive effect of the cold dark matter, although not obvious {\it a priori}, owing to the complicated field equations. We believe the constraints we have found in this work will be further verifiable in other tests of gravity and cosmology and also will be improved with the improvement of the mass versus observed size data.

One should however, also note the obvious caveat of this braneworld  model -- despite of all shifts of paradigm, we still need at least a brane cosmological constant $(\Lambda_{\sigma})$ in order to explain the observed accelerated expansion on the brane. Along with that, we have additional parameters making things much more complicated than $\Lambda{\rm CDM}$. However, as is well known, \ref{sec.1}, owing to the elusive characteristics of the dark energy/dark matter, any such viable model always warrants investigations from various perspectives. Moreover, models with extra dimensions, which should be regarded as the key thing here, can give rise to novel insight on gravity itself. It should also be noted that the brane-$\Lambda$ has at least one physical interpretation here, i.e. the brane tension.  
 
 Finally, we note that the model we have considered seems to be fully consistent  with $\Lambda{\rm CDM}$ pertaining the current cosmology from the analysis made so far~\cite{Bag:2016tvc}, also references therein with a vanishing bulk $\Lambda$. Here on the other hand, we have constrained the ``$\Omega_l-{\rm bulk}~\Lambda$'' space using the maximum turn around radius. Note that turning on an additional  source can give rise to deviation in predicted cosmological parameter(s) in any cosmological model, such as the case of the sterile neutrinos in the $\Lambda{\rm CDM}$ cosmology~\cite{Weinberg:2008zzc}. Analogously, here also in \ref{f2}, \ref{f4}, one can see that if we wish to have the value of $R_{\rm TA, max}$ on the brane to be equal to the actual observed sizes of the structures of our interest (i.e., $\sim 0.9\, R_{\Lambda{\rm CDM}}$), we can have non-zero values of $\Omega_l$ and the bulk $\Lambda$.
  However, we note that the additional sources in this case correspond to the extra dimension, thereby posing a very qualitative difference compared to models without having so. It seems important to understand whether such 
  qualitative difference could be considerable, in particular, at very large scale such as that of the Hubble horizon.    Thus, it would be interesting to extend the perturbation theory to all length scales to investigate the effect of the extra dimension on the cosmological screening mechanism (see e.g.~\cite{Eingorn:2015hza} and references therein for a discussion on the $\Lambda{\rm CDM}$). We shall return to this issue in the near future.

\section*{Acknowledgements}
\noindent
SB would like to acknowledge V.~Sahni for useful discussions on the braneworld model. SRK would like to thank K.~Patatoukos for his assistance with the graphs. The authors would like to sincerely acknowledge T.~N.~Tomaras for useful discussions and for his valuable comments on an earlier version of the manuscript. The authors would also like to thank anonymous referee for careful critical reading of the manuscript and for various useful comments.



\end{document}